# Deep Learning Mixture-of-Experts Approach for Cytotoxic Edema Assessment in Infants and Children

Henok Ghebrechristos[1*], Stence Nicholas MD [2*],
David Mirsky MD [2*], Gita Alaghband PhD [1], Manh Huynh[1], Zackary Kromer[1], Ligia Batista [2]
, Brent O'Neill MD[2], Steven Moulton MD[2], Daniel M. Lindberg MD [2]

University of Colorado Denver, Department of Computer Science and Engineering[1]
University of Colorado Denver, School of Medicine[2]

## Abstract

This paper presents a deep learning framework for image classification aimed at increasing predictive performance for Cytotoxic Edema (CE) diagnosis in infants and children. The proposed framework includes two 3D network architectures optimized to learn from two types of clinical MRI data – a trace Diffusion Weighted Image (DWI) and the calculated Apparent Diffusion Coefficient map (ADC). This work proposes a robust and novel solution based on volumetric analysis of 3D images (using pixels from time slices) and 3D convolutional neural network (CNN) models. While simple in architecture, the proposed framework shows significant quantitative results on the domain problem. We use a dataset curated from a Children's Hospital Colorado (CHCO) patient registry to report a predictive performance F1 score of 0.91 at distinguishing CE patients from children with severe neurologic injury without CE. In addition, we perform analysis of our system's output to determine the association of CE with Abusive Head Trauma (AHT) – a type of traumatic brain injury (TBI) associated with abuse – and overall functional outcome and in-hospital mortality of infants and young children. We used two clinical variables, AHT diagnosis and Functional Status Scale (FSS) score, to arrive at the conclusion that CE is highly correlated with overall outcome and that further study is needed to determine whether CE is a biomarker of AHT. With that, this paper introduces a simple yet powerful deep learning-based solution for automated CE classification. This solution also enables an in-depth analysis of progression of CE and its correlation to AHT and overall neurologic outcome, which in turn has the potential to empower experts to diagnose and mitigate AHT during early stages of a child's life.

# 1 INTRODUCTION

## 1.1 TRAUMATIC BRAIN INJURY

Traumatic Brain Injury (TBI) in children causes more than 2,000 deaths, 35,000 hospitalizations, and 470,000 emergency department visits in the US each year, making it a leading cause of pediatric disability and death [2]. The youngest children also face the additional risk of Abusive Head Trauma (AHT), a particularly deadly type of TBI that is easy to miss because caregivers rarely provide an accurate history, victims are typically pre-verbal infants, and clinical examination findings are subtle or non-specific [3]–[5].

AHT is the most important source of morbidity and mortality for abused children. The rate of AHT has been shown consistently to be 20-36/100,000 per year in the first year of life, and to be far and away the largest source of abusive mortality. Outcomes for children with AHT are dismal, with mortality rate of approximately 20% and permanent neurological sequelae in approximately 80% of survivors [6] – [9]. Victims of AHT frequently require ongoing medical support including speech, physical, and occupational therapy [5]. In the first 4 years alone, the medical costs for each case of AHT average roughly $50,000 [6].

The mechanisms and pathophysiology that differentiate AHT from other forms of TBI are poorly understood. While repetitive, rotational forces (as occur in shaking) are thought to be key to why AHT has worse outcomes compared to non-inflicted TBI (niTBI), this remains controversial, and the pathophysiological processes are not well understood [1]. Secondary injury – when damage continues after the initial trauma ends – is hypothesized to be more important in AHT than other forms of TBI, but it too is poorly understood [4], [20]. Cytotoxic Edema (CE) – a pathological process occurring when energy-requiring cell membrane ion pumps fail resulting in an abnormal influx of water into the cell leading to cell death – is thought to be central to secondary injury [20], [22] – [24].

CE has been well-described in cases of ischemic stroke and hypoxic brain injuries in adults, children, and neonates [25], [26]. AHT has been shown to produce patterns of CE similar to those seen in hypoxic-ischemic injuries, suggesting that hypoxia (either during the abusive episode itself or as a result of trauma-related apnea) is an important mediator of secondary injury [27]. CE is also thought to be more common in AHT than in other forms of TBI and to be a marker of poor prognosis [23], [28] – [30]. If CE proves to be a reliable biomarker of AHT, the problem of misdiagnosing AHT as niTBI could be mitigated in early stages. CE, if it is confirmed as specific to AHT, could also enhance understanding of the pathophysiology at work in AHT and how it differs from other forms of trauma.

No prior studies have assessed in a robust, systematic way the characteristics of CE in pediatric TBI. The proposed solution, which uses deep learning techniques to extract CE patterns form data by incorporating radiology workflow, could supply an evidence base to understand the significance of CE in pediatric TBI. More importantly, if CE provides an early predictor of neurologic outcome, this information could affect AHT treatment. Currently, the treatments of AHT are largely supportive care and monitoring for seizures [7], [8]. Better understanding of the pathophysiology of AHT, particularly the secondary injury component, which commonly occurs once a patient is under medical care where therapeutic interventions could be administered, could lead to new treatments.

We postulate that CE can help discriminate AHT from niTBI and help predict the neurologic outcome – two tasks essential to recognizing AHT and protecting the child. To that end, the development of an automated assistive tool or algorithm to classify MR-based imaging data, such as structural MRI data, and, more importantly, to distinguish subjects with features of CE from healthy subjects, can aid in not only understanding patterns of CE appearance in MRI data but also to diagnose and mitigate abuse-related head trauma in early stages. In this paper, we present a robust deep learning framework and algorithms that considers radiology workflow. The main contributions of this work can be listed as follows:

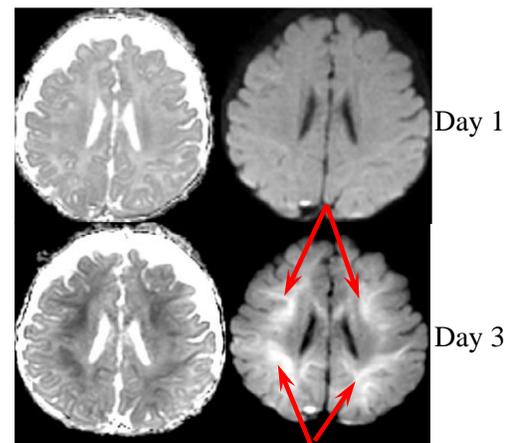

Day 1

Day 3

**Figure 1.** As cells swell due to inward shift of water, there is a commensurate decrease in diffusion, identified as high signal on DWI (right) and low signal on ADC (left).

- We present a novel deep learning framework and network architectures optimized to learn patters of Cytotoxic Edema from volumetric time slices of brain MRI of infants and children.
- We evaluate the performance of the proposed solution and present results that demonstrate the efficacy of the framework in predicting the presence or absence of Cytotoxic Edema using clinical data of infants and children less than 10 years of age admitted to Children's Hospital Colorado (CHCO) [9].

## 1.2 CYTOTOXIC EDEMA IMAGE CHARACTERISTICS

Cytotoxic edema (CE), a type of cerebral edema, is the result of cells being unable to maintain ATP-dependent sodium/potassium (Na+/K+) membrane pumps which are responsible for high extracellular and low intracellular Na+ concentration [10]. When an insult to the brain results in ischemia or hypoxia, this oxygen-requiring process produces very little to no new ATP. Cells quickly use up their reserves of ATP and, unless normoxia (normal oxygen level) is restored, the cellular machinery loses its ability to sustain homeostasis. This results in intracellular swelling and reduction in the extracellular volume, the combination of which causes a reduction in the observed degree of isotropic water diffusion. The trace DWI is generated by summing the signal of water diffusion in 3 perpendicular directions in space. This trace DWI image displays areas of abnormally reduced water diffusion as hyperintense (bright), but it is also T2 weighted and thus can be confounded by areas that are also T2 hyperintense. Therefore, an ADC map is usually also automatically created by the scanner. This is a parametric map of the signal intensity of a purely T2* weighted image subtracted from the trace DWI. Voxels in this ADC map therefore represent the calculated, actual ADC values at a given voxel, which by convention are become darker with greater degrees of diffusion restriction. Thus, areas of restricted diffusion indicating CE are represented on imaging as bright or hyperintense on DWI, and dark or hypointense on ADC (Figure 1). For details on clinical and imaging characteristics of CE, refer to publications by Liang et. al. [11] and Rosenblum [10].

The rest of the paper is organized as follows:

- We first present related work and relevant research outcomes on the domain problem;
- We then present the method of study where we detail study population, dataset preparation and system design;
- We then follow that section by sharing our results and discussion before we present conclusion remarks and future direction of this research.

## 2 RELATED WORKS

Advances in Magnetic Resonance Imaging (MRI) enabled the non-invasive visualization of the infant's brain through acquired high-resolution images [12], [13]. The increasing availability of large-scale datasets of detailed infant brain multi-modal MR images (e.g., T1-weighted (T1w), T2-weighted (T2w), diffusion-weighted MRI (dwMRI), and resting-state functional MRI (rsfMRI) images) affords unique opportunities to accurately study early Abusive Head Trauma (AHT), leading to insights into the origins and abnormal developmental trajectories of CE.

Changes in brain structure and function caused by CE and their correlation to TBI, morbidity, and overall outcome have proved of great interest to research groups. In diagnostic imaging of adults in particular, machine learning based classification and predictive modeling of the stages of Cerebral Edema [14] in large cohorts of TBI patients have been investigated [15]– [18].

Amorim et. Al. [17] investigate machine learning models for the predictive performance of mortality and length of stay applied to a low- and middle-income countries (LMIC) cohort of TBI patients. They explore various machine learning techniques using predictors such as gender, age, and level of pupil reactivity at admission and report a high prediction performance, with the best prediction for overall mortality achieved through Naive Bayes (area under the curve (ROC) = 0.906).

Hale et. al. [15] build and validate an artificial neural network (ANN) using a prospectively collected, publicly available, multicenter TBI dataset. They use clinical and radiologist-interpreted imaging metrics in order to predict Clinically Relevant TBI (CRTBI). Among 12,902 patients included in this study, 480 were diagnosed with CRTBI. The authors' ANN had a sensitivity of 99.73% with precision of 98.19%, accuracy of 97.98%, negative predictive value of 91.23%, false-negative rate of 0.0027%, and specificity for CRTBI of 60.47%. The area under the receiver operating characteristic curve was 0.9907.

In contrast to previous work, our study is the first of its kind. It lays the groundwork for future studies incorporating brain edema, particularly Cytotoxic Edema in infants and young children. Based on guidance from domain experts, the proposed solution considers the manual procedures and radiology workflow. We design a system that uses weighted average of two distinct models; ADCNet and DWINet, each tuned to learn patterns of CE from ADC and DWI maps, respectively. This approach achieves radiology-level predictive performance at differentiating brain scans with CE from those that are normal.

# 3 METHODS

## 3.1 STUDY DESIGN AND POPULATION

This study is a prospective analysis of MRI data acquired in patients with Abusive Head Trauma (AHT). The study was approved by the institutional review boards of University of Colorado and Children's Hospital Colorado (CHCO), and all experiments were performed in accordance with relevant guidelines and regulations. All data were anonymized, and this study was fully compliant with the Health Insurance Portability and Accountability Act.

**Table 1.** Summary of Clinical Data for the TBI Cohort.

| Total MRI Records | Distinct Patients | Gender (male) | Number of Patients with 2 or more MRIs | DWI \ ADC | CE + \ - | Abuse + \ - |
|---|---|---|---|---|---|---|
| 463 | 291 | 65% | 38% | 287\ 286 | 152 \ 136 | 196/93 |

Potential patients were identified using existing records in the CHCO trauma registry – a prospectively maintained database mandated by the state of Colorado of all trauma centers. All trauma patients seen at CHCO are entered into the database nearly concurrent with their hospitalization. The patient population identified by the trauma registry were cross-referenced to the radiology database and the child protection team database to ensure completeness of subject identification.

| FSS Score Range | Outcome Assessment | Count |
|---|---|---|
| <= 6 | Good (no disability) | 158 |
| (6, 14] | Mild disability | 58 |
| (14, 21] | Moderate disability | 59 |
| >= 21 | Sever & vegetative disability | 8 |

**Table 2.** Categories of baseline functional status assessment according to baseline PGOS-E5 and FSS6 ratings assigned 6-12 months after the initial injury.

Masked review of MR images was used to determine the presence, extent, and distribution of CE, and inpatient and outpatient charts review to determine the cause of TBI (abusive/non-inflicted) and disability as defined by in-hospital death, PGOS-E5, and FSS6 - both validated outcomes status scales that can be reliably determined by retrospective review. Demographic, injury, and treatment data was also collected. See Tables 1 and 2 for summaries of dataset.

All data was collected by trained study personnel and stored in a secure REDCap database on the University of Colorado Denver server. In this feasibility study, we included MR images from the baseline examination and follow-ups up to 6 months. The initial dataset included 463 scans with 291 unique subjects each having one or more exam sessions. Two of the total MRIs were corrupt and unusable.

To combat overfitting resulting from data scarcity, we also used the publicly available adult brain MRI, IXI dataset maintained by the Imperial College of London [40]. This dataset is used for gender classification and is the best publicly available dataset we could find to initialize our models, battle overfitting, and expedite training.

### 3.1.1 Input Data

All input images to the learning system were captured using Diffusion-Weighted Imaging (DWI) [20] modality. The raw DWI map along with the Apparent Diffusion Coefficient map – a quantitative evaluation of the diffusion image (Figure 1) – were used to train and validate an ensemble of deep neural networks.

DWI evaluates the molecular function and micro-architecture of the human brain. DWI signal contrast can be quantified by ADC maps, and it acts as a tool for treatment response evaluation and assessment of CE progression. Diffusion is qualitatively evaluated on trace images and quantitatively by the parameter called ADC. Tissues with restricted diffusion are bright on the trace image and hypointense on the ADC map (Figure 1, left).

### 3.1.2 Dataset Preparation

Each patient often has multiple MRI series performed as part of the first and follow-up sessions. Sampling strategy detailed in Figure 2 was employed to ensure MRI sample sets from the two consecutive sessions (baseline and follow-up) of the same subject were assigned to separate groups – either training or testing. For training the Deep Learning (DL) classification system, labels were generated for each MRI volume and subject based on the presence or absence of CE by expert radiologist in a double-blind review process. In addition to CE labels, labels indicating abuse diagnosis and Functional Status Scale (FSS) score for each subject were used to train DL models. Summary statistics of the dataset is presented in Tables 1 and 2.

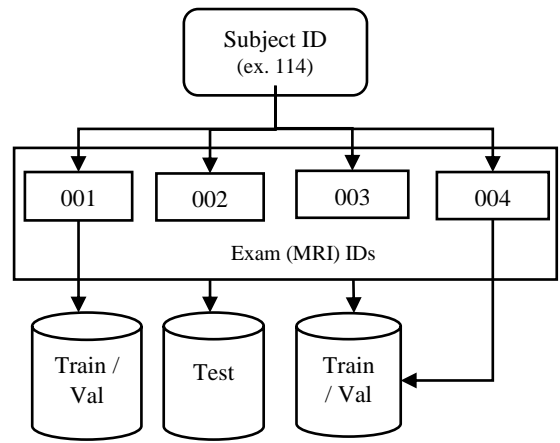

**Figure 2.** Sampling strategy.

#### 3.1.2.1 Preprocessing
The data were augmented by including 90° rotations and reflections about XY and YZ planes to avoid overfitting during network training. Non-brain regions, including skull and neck voxels, were removed. Further processing was needed to exclude series that fall outside the nominal dimension of the voxel region common across the dataset. Each volume was preprocessed so that the brain voxel fits within a fixed 1.5mm isotropic window. Model-specific preprocessing outcomes are highlighted in the results section of this article.

### 3.1.3 CyteNet - Machine Learning Framework

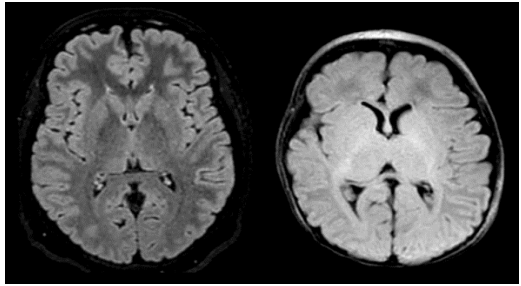

**Figure 3.** Comparison between Adult (left) and Child (right) MRI scans of the brain.

The analysis of MR infant brain images is typically far more challenging as compared to the adult brain setting. As illustrated in Figure 3, an infant's brain MRI suffers from reduced tissue contrast, large within-tissue inhomogeneities, regionally heterogeneous image appearance, large age-related intensity changes, and severe partial volume effect due to the small brain size. Since most of the existing machine learning tools were designed for 2D image data and for adult brain MRI, infant-specific machine learning framework (CytENet) and custom 3D network architecture were designed to learn abnormalities from 3D infant volumetric MRIs.

We recast the detection of CE into a binary classification problem and used an ensemble approach to learn and identify patterns of CE. We employed custom and ultramodern convolutional neural network architectures having distinctive characteristics to optimize learning and generalization performance. The weighted average ensemble model consisted of two networks. To account for functional and structural dissimilarities between how CE effects show on DWI and ADC maps, separate preprocessing and feature learning pipelines were implemented (Figure 5). Inspired by radiologists' workflow to image analysis for diagnosis, the first network – DWINet was designed to extract features from DWI maps, while a second network, ADCNet, with ResNet18 backbone was used to learn patterns of CE from ADC maps as described next.

The design of both model architectures was informed partly by failed trial and error experiments using off-the-shelf model architectures designed for image classification. For DWI path, we first tried variants of ResNet [21] and EfficientNet [22] model architectures as backbones for binary classification of CE+ \ - using the DWI maps to no avail. Both models converged during training but were unable to achieve reasonable validation performance. We tried warm initialization and training from scratch. In both cases, we saw validation accuracy stagnation below 55%. This pointed at the models' inability to learn and effectively generalize learning to unseen samples. We believe these performance caps are attributed to two reasons. First, these model architectures are designed for large amounts of data and common image formats (not medical images), requiring training from hours to days using extensive number of images per category. In addition, we believe DWI image characteristics is a culprit for the low performance. Since most of the DWI images were captured with high diffusion gradient strength or b-value, they contain high noise levels that overwhelm the signal of interest (Figure 1, right). This makes the models susceptible to learning noise rather than the signals showing CE presence. We considered preprocessing denoising techniques [23] design for DWI images. We saw more stability and higher validation performance during training. However, the generalization performance was not on par for clinical usage. To overcome these limitations, we experimented with several custom network architectures varying in depth of convolution layers and kernel size and select the most stable model. The design process considered DWI image characteristics and pixel distribution of the DWI maps. Ultimately, we found that a smaller network with a stack of convolutional layers that apply consecutive 3 by 3 kernels revealed the best results.

In contrast to DWI map, ADC maps are high contrast images and do not suffer from high noise (low signal) issue (Figure 1, left). To learn patterns of CE from ADC, we used ResNet18 backbone with binary classification output. Each map is first converted into 3-channel image by stacking each monochrome slice to form an RGB image. This approach resulted in high performance during training and generalization performance during testing. The results are discussed in section 4. Detail of each model architecture is presented below.

**DWINet** (Figure 4B) – the champion DWINet model consists of three 3D convolutional blocks and one block of fully connected layers followed by a softmax activation function at the output. Each convolutional block is composed of multiple 3D convolutional layers, each with rectified linear unit activation, followed by a batch normalization and max pooling layer (Figure 4B). Convolutional layers on each block are assigned a specific number of filters, beginning with 64 for convolutional layers in the first convolutional block. Subsequent blocks have double the number of filters from the previous block. A global average pooling is applied to the output of the convolutional blocks before they are passed to the fully connected network. In our implementation, the receptive field (convolution kernel) was set at 3 3 3 with a stride size of 1 for all convolutional layers. The second fully connected network, which combined the volume scores for subject-wise prediction, consists of a classification network, a hidden layer of 256 nodes, and an output layer with a sigmoid activation function.

**ADCNet** (Figure 4C) – is a 3D convolutional network adopted based on the ResNet-18 architecture [24], which represents a good trade-off between depth and performance. It includes five convolutional stages (see Figure 4C for detail). The first convolutional layer applies a 7 by 7 convolution kernel at stride 2 and produces 64 feature maps. The first convolution stage is followed by a max pooling layer that returns a down-sampled output by applying a 3 by 3 window at stride 2. The subsequent convolution stages apply two 3 by 3 convolutions. In the second stage the number of filters is kept the same at 64, the subsequent stages use double the number of filters of the previous stage. The convolved outputs are then passed through an average pooling operator to produce features. A fully connected network with 512 nodes followed by an output layer with softmax activation function is attached to the end of the network to classify the features into negative or positive outcomes.

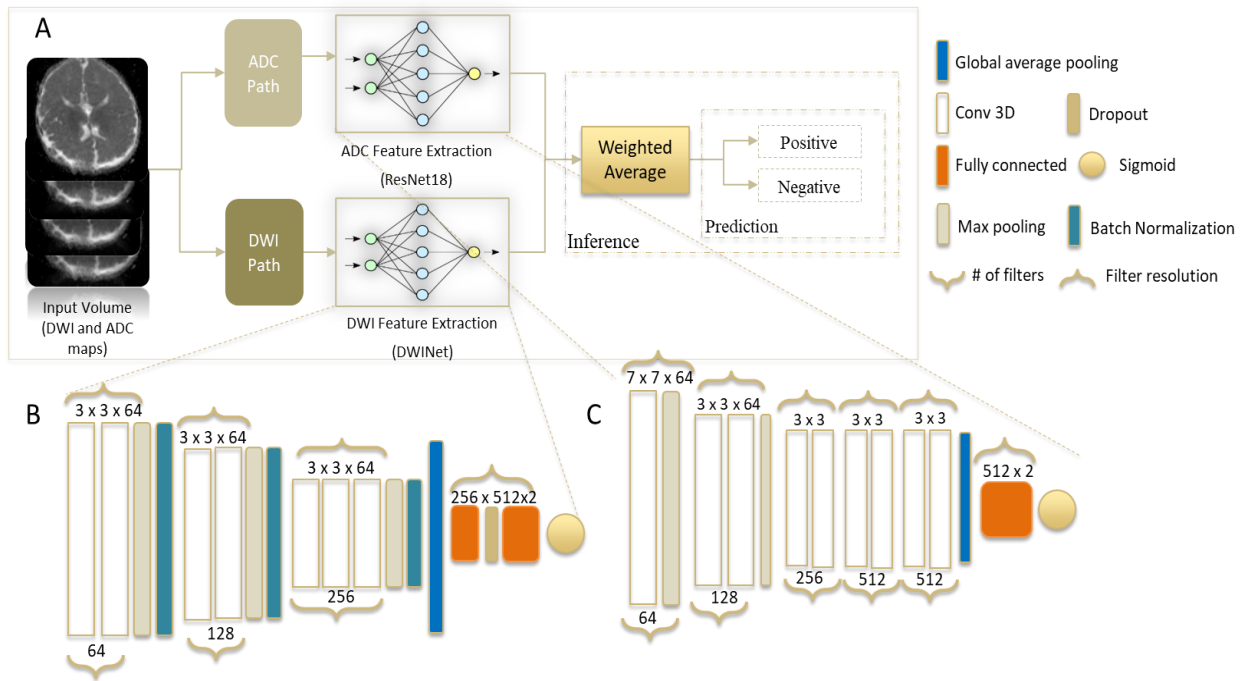

**Figure 4.** A, Illustration shows design of weighted average system architecture for predicting Cytotoxic Edema. B, Illustration shows DWINet network architecture that extracts features from DWI maps. C, Illustration shows ADCNet with ResNet18 backbone for learning patterns of CE from only ADC maps. Scores from each network were combined to produce overall volume prediction. Numbers above each layer in B and C indicate image resolution; numbers underneath layers indicate number of filters (kernel size, 3 3 3).

**Network Training** – the training procedure, which was modeled after the manual radiology process of analyzing MRI scans, consisted of two input pipelines: DWI and ADC. The DWI path trained the DWI network to produce and classify features from normal and abnormal scans. The training set used in this stage consisted of only DWI maps. A dataset consisting only ADC maps was fed to the ADCNet through the ADC path detailed in Figure 4. ADCNet generates and classifies features from ADC maps belonging to CE positive or negative classes. The outputs from the two networks were then passed through an average weighted network to optimize generalization performance. We implemented this procedure based on guidance from experts of the domain. The manual effort involved in identifying CE from MRI scans usually requires visual analysis of both maps. Each map serves a different purpose. The DWI map is generally used to identify regions of interest while the ADC map, which captures the strength of diffusion, is used to do an in-depth analysis. In both setups, Adam, an algorithm for first order gradient-based optimization of stochastic objective functions, was used as an optimizer [25] owing to its fast convergence and weight dependent learning rate. Binary cross-entropy was used as the model loss function along with rectified linear unit activation for DWINet and sigmoid for ADCNet. In the final stage, the last convolutional block and the dense layers were trained on the image features.

## 4 EXPERIMENTAL RESULTS

The area under the receiver operating characteristic curve (AUC), sensitivity, specificity, and accuracy were calculated to evaluate each network's performance. This analysis was performed for subject-wise predictions. A three-fold cross-validation procedure was implemented to assess the stability of the DL model during training. The data were split into two major sets for model development: of 80% of the data set (65%) was used for training and 15% for validation. The remaining 20% was held out and used for testing. This process was repeated three times by changing the assignments for the partitioned training data. All processing was performed on Heracles multi-core cluster of the Department of Computer Science and Engineering's Parallel Distributed Systems (PDS) Lab at the University of Colorado Denver. The cluster consisted of 18 nodes, one master, 16 compute nodes and a GPU node with four graphics processing units (Tesla V100; Nvidia). Implementation was carried out by using the Keras Python library (version 2.4) [26] and TensorFlow (version 2.8.0) [27].

### 4.1 MODEL PERFORMANCE

Each model was initially trained using the IXI dataset [19] before it was fine-tuned on the custom dataset. We trained each model for a minimum of 300 epochs using a cross-entropy [28] loss function. Validation during training was set to run every 100 steps. The trends observed during model development are presented in Figure 5.

The model's architecture allows sharp learning during the initial epochs and a steadier trend throughout the rest of the epochs (Figure 5). Furthermore, hyperparameters tuning was not used to train the model and that could explain the fluctuation and spikes of the results on the training set. However, the models gave comparable results on the validation partition and on the test partition (unseen images) in terms of accuracy and macro F1 score. An average training accuracy of 93% for DWINet and 99% for ADCNet was obtained using the framework and training pipeline detailed in Section 3. In a three-fold validation, DWINet achieves an average 98% validation accuracy separating CE positive scans from normal scans, while ADCNet achieves 96% accuracy.

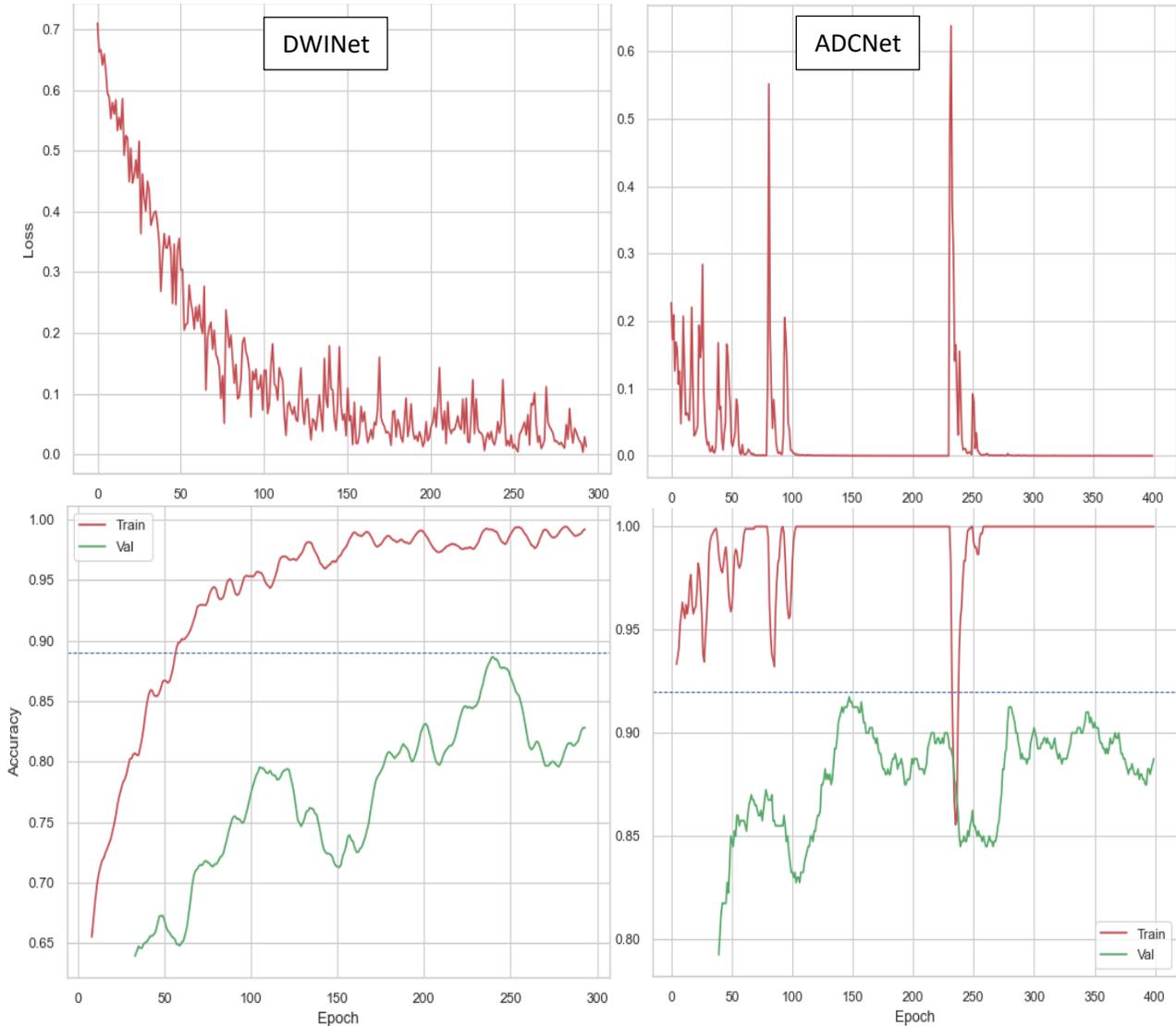

**Figure 5.** Graphs show training performance observed during model development. Subject wise three-fold cross-validation was performed for DWINet (Left) and ADCNet (Right). The loss (Top) and accuracy (bottom) trends achieved by both networks are consistent with models that are learning distinguishing features from the underlaying dataset.

### 4.1.1 Subject-level Recognition

To expand the application for distinguishing Cytotoxic Edema subjects from normal healthy brains to the clinical level, subject-level generalization of our solution was investigated. The receiver operating characteristic (ROC) curves of top candidate models are illustrated in Figure 6 (right). Across all candidate models for ADC and DWI, the results of the top two (candidate 1 of ADC and candidate 3 of DWI) and their weighted average is presented in Table 3.

To obtain subject level diagnosis, slice level decisions at different class probability thresholds varying in the range of [0,1] were considered. The corresponding areas under the curve (AUC) scores of three top candidates at prediction thresholds of 40, 50, and 60 percent were compared (Figure 6, right). Majority voting of slice level predictions is used to determine subject level classification. As seen in Figure 6, the model achieves the highest AUC of 0.99 (candidate 3, Figure 6 right) and 0.97 (candidate 1, Figure 6 right) for DWI and ADC respectively. The testing accuracies when using these two models are 92% and 80%, respectively. The best patient level AUC achieved during validation reaches 0.97 (left plot on Figure 6). The overall performance of the system is presented under weighted average row of Table 3. The combined score of the proposed solution on the test dataset (AUC column of Table 3) comfortably exceeds that of the baseline on the validation set. These performance metrics confirm that the performance of classifiers inclined toward the ideal diagnostic performance and were beyond random guess.

| Model | Precision | Recall | Specificity | Sensitivity | F1-Score | AUC | ACC (%) |
|---|---|---|---|---|---|---|---|
| ADC Network | 0.89 | 0.89 | 0.78 | 0.75 | 0.60 | 0.97 | 80 |
| DWI Network | 0.93 | 0.90 | 0.84 | 0.97 | 0.91 | 0.99 | 92 |
| **Weighted Average** | **0.91** | **0.83** | **0.81** | **0.86** | **0.755** | **0.98** | **86** |

**Table 3.** Generalization performance results of our weighted average ensemble system on test data for the Prediction of Cytotoxic Edema.

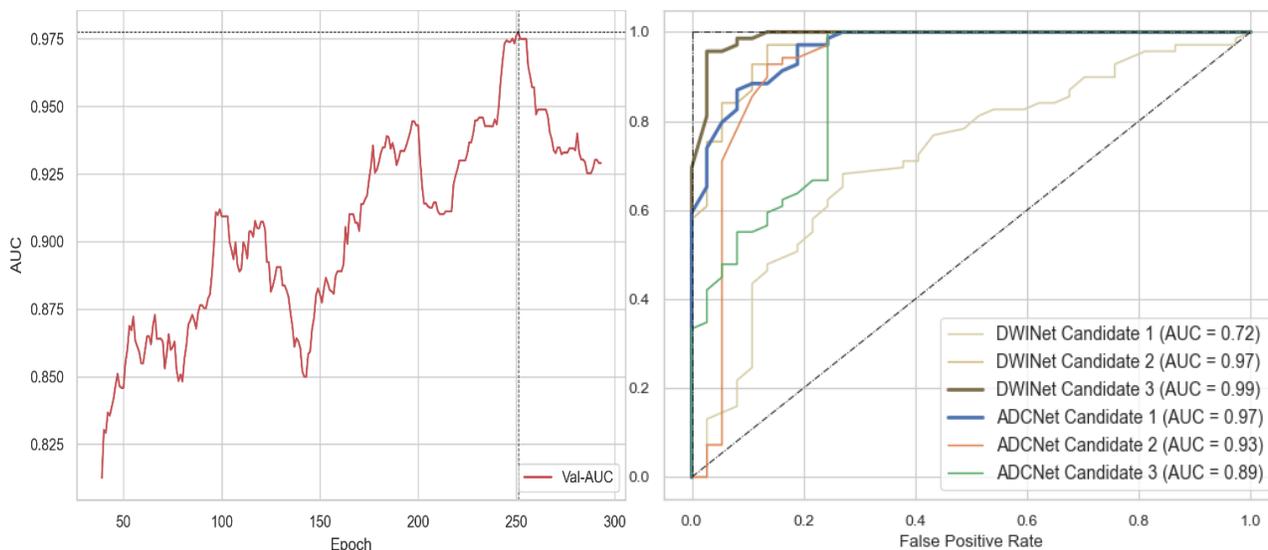

**Figure 6.** Graphs show receiver operating characteristic (ROC) curves (Right) of several candidates with true-positive rate plotted against false-positive rate. Dashed line is the chance line. Thin lines show receiver operating characteristic curves for three runs each producing slice wise prediction with different probability thresholds – 40 %, 50%, and 60% for candidate 1, 2 and 3 respectively.

## 4.2 CE CORRELATION TO AHT AND NEUROLOGIC OUTCOME

Identification of CE in its early stages is crucial to mitigate AHT. Here we present the correlational results of CE to AHT and overall functional outcome (in-hospital child morbidity and mortality). To understand how well the presence of CE predicts AHT and overall outcome, we used logistic regression of CE predictions against AHT diagnosis and Functional Status Scale (FSS) score. Summary statistics of the dataset used for this is presented in Table 2. The results are depicted in Figure 7.

**CE correlation to AHT** - to determine if CE can be established as a biomarker of AHT and to examine correlation between extent of CE and extent of disability from AHT, we applied binary logistic regression. For this analysis, the dependent variable (AHT) was transformed into dichotomous category; AHT + and AHT –, based on whether the source to the subject's TBI is rooted in Abuse or not. We discover that the proposed solution produces $R^2$ value of 0.5 when associating probabilities of CE prediction to Abuse and non-Abuse diagnosis (Figure 7, top).

**CE correlation to overall outcome** - establishing CE as a predictor of long-term outcome has even greater potential impact on the care of TBI patients. We used Functional Status Scale (FSS) score as a measure of morbidity and overall functional outcome. The logistic regression model was run on the probabilities of CE presence in each scan against the discrete variable FSS score. We found a statistical significance correlation with $R^2$ value 0.64 (Figure 8, bottom).

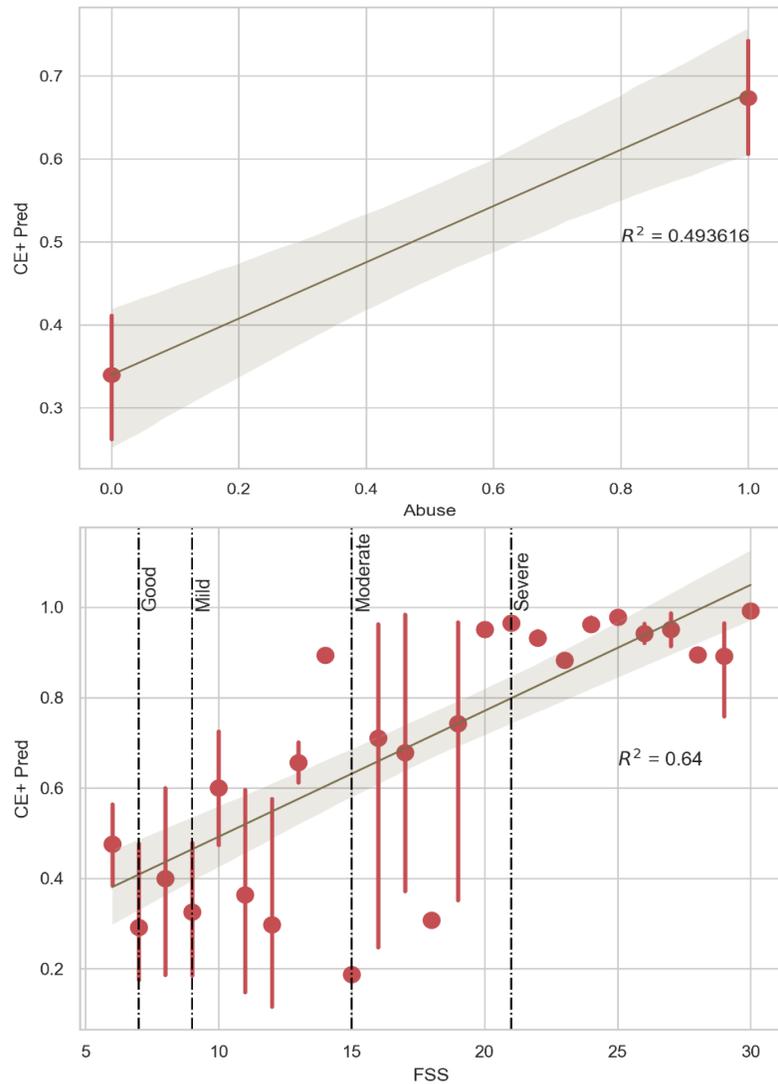

**Figure 7**. Cytotoxic Edema predictions correlates with functional assessment score of each subject (Highlighted zone (Bottom) represents 64% confidence interval for predictions from the logistic regression model). Correlation based on solely logistic regression applied to AHT labels is inconclusive. (Highlighted zone (Top) represents only 50% confidence of associating CE predictions with Abusive or non-abusive labels.

Discrimination of CE + scans were high for all models, while logistic regression showed the numerically highest discrimination in associating CE predictions to overall outcome and morbidity. Further study and additional variables are required to determine how well CE is correlated to abuse-related TBI.

# 5 CONCLUSION

This study provides evidence that we can automate brain imaging data analysis and obtain meaningful results on small cohorts of AHT patients. Such an approach – leveraging routinely obtained clinical imaging data or imaging obtained in clinical trials to advance the science brain edema – is the pathway to realizing the potential of artificial intelligence in brain imaging of young children. We presented two robust training pipelines that use deep learning-based classifiers of MRI to distinguish brains affected by CE and healthy brains in infants and children. Scale- and shift-invariant low- to high-level features were extracted from DWI and ADC maps using convolutional neural network architecture, resulting in a significantly accurate and reproducible predictive ensemble model. In this study, the achieved accuracy rates are comparable to clinicians' performance. Furthermore, two types of MRI data (DWI and ADC) of subjects of age 10 years or less and clinically diagnosed with CE were used to train a deep learning-based system for the first time. This successful and innovative deep learning-based framework has implications for numerous applications in classifying brain disorders in both clinical trials and large-scale research studies. This study also demonstrated that the developed separate pipelines served as fruitful approach in characterizing multimodal MRI. Lastly, the subject-level classification designed for clinical purposes enabled us to assess the robustness of the MRI pipelines followed by a decision-making algorithm, verifying that they both distinguish Cytotoxic Edema patients from those with severe brain injury but without CE with high accuracy. Discrimination was high for all models, while logistic regression showed CE and AHT had low correlation, while CE and child morbidity and mortality had statistically significant correlation.

Further study and incorporation of more clinical data and variables is necessary to conclude whether CE is a biomarker of AHT. We believe our work paves the way to further study the stated hypothesis in detail.

# 6 FUNDING

This work was supported by the Colorado TBI Trust Fund (MINDSOURCE).